\documentclass[aps,prd,showkeys,showpacs,twocolumn,a4paper]{revtex4-1}

\usepackage{ulem}
\usepackage{color}
\usepackage{graphicx}
\usepackage{appendix}
\usepackage{epsfig}
\usepackage[colorlinks=true,linkcolor=blue]{hyperref}
\usepackage{epstopdf}
\usepackage{amsmath}
\usepackage{subfig}
\usepackage{comment}
\usepackage{float}
\usepackage[dvipsnames]{xcolor}
\newcommand{\be}{\begin{equation}}
\newcommand{\ee}{\end{equation}}
\newcommand{\ba}{\begin{eqnarray}}
\newcommand{\ea}{\end{eqnarray}}

\usepackage[cp1252]{inputenc}
\usepackage{eufrak}

\begin{document}
\title{Viscous QCD medium effects on the bottom quark transport coefficients}

\author{Adiba Shaikh}
\email{adibashaikh9@gmail.com}
\affiliation{Department of Physics, Indian Institute of Technology Bombay, Powai, Mumbai-400076, India}

\author{Sadhana Dash}
\affiliation{Department of Physics, Indian Institute of Technology Bombay, Powai, Mumbai-400076, India}

\author{Basanta K. Nandi}
\affiliation{Department of Physics, Indian Institute of Technology Bombay, Powai, Mumbai-400076, India}

\begin{abstract}

The bottom quark transport coefficients, i.e., drag and diffusion coefficients, have been studied for the collisional and soft gluon radiative processes within the viscous QCD medium. The thermal medium effects are incorporated using the effective fugacity quasiparticle model (EQPM). Both the shear and bulk viscous effects at leading order are embedded through the near-equilibrium distribution functions of the quark-gluon plasma (QGP) constituent quasiparticles. The transport coefficients' dependence on the bottom quark's initial momentum and QGP temperature has been investigated. The relative dominance of the radiative over the collisional process for the bottom quark seems to occur at a higher initial momentum compared to that of the charm quark. In contrast, the effect of the viscous corrections seems to be more for the charm quark. 

\end{abstract}


\keywords{Bottom quark, Drag and diffusion, Energy loss,  Shear and bulk viscous correction.}

\maketitle

 \section{Introduction}

Hadrons, under extreme conditions like high temperatures ($T \gtrsim 200$ MeV $\approx 10^{12}\, K$) and/or high baryonic densities ($\mu \gtrsim 200 $ MeV) undergo phase transition into the deconfined state of quarks, anti-quarks and gluons as the effective degrees of freedom. Such conditions which are believed to exist only in the early universe and inside the core of the neutron star are recreated in the experiments at some of the largest particle accelerators like the Relativistic Heavy Ion Collider (RHIC) and Large Hadron Collider (LHC) in a controlled environment to investigate its properties ~\cite{Adams:2005dq,Back:2004je,Arsene:2004fa,Adcox:2004mh}. Relativistic viscous hydrodynamics have successfully described the evolution of the QGP phase involving various dissipative processes~\cite{Heinz:2013th,Gale:2013da, Jaiswal:2016hex}. The transport coefficients which are sensitive to medium evolution are theoretically calculated by the underlying microscopic theory like the effective kinetic theory approach ~\cite{Aamodt:2010pb,Jaiswal:2020hvk} and compared with those extracted experimentally. Within dissipative hydrodynamics, the effect and evolution of the QGP medium through the spatial anisotropy and pressure gradients are explored by introducing the shear and bulk viscosity respectively ~\cite{Romatschke:2007mq,Ryu:2015vwa}. \\

Heavy quarks like charm and bottom are created in the early stages of the heavy-ion collision, primarily due to partonic hard scattering such as the gluon fusion process. Such heavy quarks within the medium introduce an energy scale, i.e., their mass ($m_{HQ}$) which is an order of magnitude larger than the temperature of the QGP medium ($T\approx500$ MeV). Due to their large mass ($m_c \approx  1.3$ GeV, $m_b \approx  4.2$ GeV), they do not thermalize with the plasma over its lifetime and traverse through the plasma unequilibrated. Hence, they act as an excellent probe to study QGP evolution and its properties 
\cite{Andronic:2015wma, Prino:2016cni, Aarts:2016hap, Rapp:2018qla, Dong:2019unq}. Also, heavy quarks, being heavier than the typical strong interaction confinement scale $\Lambda_{QCD}\approx200$ MeV, can be treated non-relativistically having a small strong coupling constant ($\alpha_s\approx 0.1-0.3$) in the realm of perturbative QCD. Heavy quarks undergo energy loss while traversing through QGP due to collision (2 $\rightarrow$ 2, elastic) with the plasma constituents and radiation of the soft gluons (2 $\rightarrow$ 3, inelastic). Collisional energy loss is dominant for the heavy quark with low momentum whereas at high momentum medium induced gluon radiation becomes dominant ~\cite{Braaten:1991we,Mustafa:2004dr,Mazumder:2011nj,Abir:2012pu,Cao:2013ita,Sarkar:2018erq,Cao:2016gvr,Zigic:2018ovr,Liu:2020dlt,Uphoff:2014hza,Gossiaux:2006yu,Das:2010tj,Cao:2015hia}. \\

The movement of heavy quarks within QCD plasma can be treated as a Brownian motion of massive particles within the fluid. Hence, the phase space evolution of the heavy quarks is governed by the Boltzmann transport equation. Further, the soft-scattering approximation of momentum transfer between heavy quarks and in-medium particles reduces the transport equation to the Fokker-Planck equation, where its interaction with the medium constituents (quarks, anti-quarks and gluons) is incorporated through the drag and momentum diffusion coefficients ~\cite{Svetitsky:1987gq, GolamMustafa:1997id, Das:2009vy}. Heavy quark transport coefficients are sensitive to the medium evolution in the presence of dissipative processes within QGP. The transport coefficients associated with the dissipative processes (viscosity, electric conductivity, etc.) in the hot and dense QCD medium can be determined from the underlying microscopic theories, such as the effective kinetic theory approach. They could also be extracted experimentally at RHIC and LHC through various observables and phenomenological transport models and, using lattice QCD. ~\cite{He:2022ywp, ALICE:2021rxa,Cao:2018ews,Das:2013kea,vanHees:2005wb,Gossiaux:2008jv,Xu:2018gux,Das:2015ana,Scardina:2017ipo,Song:2015sfa,Adler:2005xv,Adare:2006nq,Alberico:2013bza,vanHees:2007me,Li:2019wri,Jamal:2021btg,Akamatsu:2008ge,Uphoff:2011ad,Xu:2013uza,Banerjee:2011ra,Brambilla:2020siz}. The viscous corrections to the heavy quark transport coefficients for the collisional processes have been previously studied ~\cite{Das:2012ck, Song:2019cqz,Singh:2019cwi, Kurian:2020kct, Kurian:2020orp}. Here, we go a step further to investigate the effect of viscous corrections on the radiative process of the bottom quarks as a follow-up to our previous study for the charm quark ~\cite{Shaikh:2021els}. \\

In this work, we study the impact of the shear and bulk viscosities of the QGP on the drag and diffusion coefficients of the bottom quark for collision and radiative processes. The thermal QCD medium interactions with the realistic equation of state effects are incorporated using the effective fugacity quasiparticle model (EQPM) ~\cite{Chandra:2011en,Chandra:2008kz}. This model considers QGP as a medium of non-interacting quasiparticles with dynamical effects included in the dispersion relation of constituent particles through the introduction of the effective fugacity parameter. For viscous corrections, the non-equilibrium distribution function has been obtained by solving the effective Boltzmann equation based on the EQPM with the relaxation time approximation (RTA) and using the Chapman-Enskog like iterative expansion method~\cite{Mitra:2018akk}. The collective mean-field contributions originating from the underlying conservation laws are included in the determination of the near-equilibrium momentum distribution functions. Here, we present the result for the effect of shear and bulk viscous corrections on the radiative process of the bottom quark in terms of its transport coefficients and compare it to the collisional process. We observe a considerable modification in the drag and diffusion coefficients of the bottom quark due to these corrections, compared to the charm quark previously studied in ref. ~\cite{Shaikh:2021lka}. The bottom quark is roughly three times heavier than the charm quark with a large thermal relaxation time and intrinsic mass scale being very far from the non-perturbative QCD confinement scale ($\Lambda_{QCD}\approx200$ MeV). Therefore, it is a better probe for the investigation of the in-medium properties via its transport coefficients and motivation for the present work. \\
 
 In section \ref{Formalism}, the formulation of the heavy quark dynamics in the QCD medium is discussed for collisional and radiative processes followed by the EQPM description of viscous corrections to the momentum distribution function of the quasiparticles. Section \ref{Results} focuses on the results for the transport coefficients of the bottom quark including shear and bulk viscous corrections followed by the comparative study of the energy loss of charm and bottom. We summarize with the conclusion in section \ref{Conclusion}.\\

\noindent {\bf Notations and conventions}: The subscript $k$ denotes the particle species, where $k=lq$ represents light quarks, $k=l\bar{q}$ represents light anti-quarks and $k=g$ representing the gluons. The degeneracy factor for gluon is $\gamma_g=N_s\times (N_c^2-1)$ and for light quark (antiquark) is $\gamma_{lq}=N_s\times N_c\times N_f$ with $N_s=2$, $N_f=3$ ($u,d,s$), and $N_c=3$ (for $SU(3)$). $u^{\mu}$ is the normalized fluid velocity satisfying the relation $u^{\mu}u_{\mu}=1$ with the metric tensor $g^{\mu\nu}=\text{diag}(1, -1, -1, -1)$.

\section{Formalism}\label{Formalism}

%
\subsection{Heavy quark transport coefficients}
%
We can treat the bottom quark as a Brownian particle (non-equilibrated) moving within the QGP medium (equilibrated) where the massive quark loses its energy due to collision with the medium constituents (elastic process) and also through gluon radiation (inelastic). Both these interactions of the bottom quarks within the QGP medium are incorporated in their transport coefficients. 
%
\subsubsection{\textbf{\textit{2$\rightarrow$2 process}}}
%
The bottom quark $(HQ)$ undergoes collisions with the medium constituents, $i.e.$, light quarks $(lq)$, light antiquarks $({\bar lq})$ and gluons {$(g)$}. The elastic ($2\rightarrow 2$) process is,
\begin{equation}
    HQ \ (p)+lq/l\bar{q}/g \ (q) \rightarrow HQ \ (p')+lq/l\bar{q}/g  \ (q').\label{Col}
\end{equation}  

The evolution of the bottom quark momentum distribution function $f_{HQ}$ is determined by the Boltzmann transport equation, which within the soft scattering approximation reduces to the Fokker-Planck equation ~\cite{Svetitsky:1987gq}.
\begin{align}\label{1.1}
  	\frac{\partial f_{HQ}}{\partial t}=\frac{\partial}{\partial p_i}\left[A_i({\bf p})f_{HQ}+\frac{\partial}{\partial p_j}\Big(B_{i j}({\bf p})f_{HQ}\Big)\right],
\end{align}
where $A_i({\bf p})$ is the drag force $B_{ij}({\bf p})$  is the momentum diffusion tensor for the bottom quark expressed in the form of thermal average as,
\begin{align}
   A_i({\bf p}) =& \,\frac{1}{2 E_p  \gamma_{HQ}} \int \frac{d^3 {\bf q}}{(2 \pi)^3 E_q}  \int \frac{d^3 {\bf q'}}{(2 \pi)^3 E_{q'}} \int \frac{d^3 {\bf p'}}{(2 \pi)^3 E_{p'}}\nonumber\\
   &\times\sum{|{\mathcal{M}}_{2\rightarrow2}|^2} \ (2 \pi)^4 \  \delta^{(4)}(p+q-p'-q') \nonumber\\
   & \times f_k(E_q) \  (1\pm f_k(E_{q'})) \ [(p-p')_i]\nonumber\\
   =& \ \langle\langle ( p - p')_i \rangle\rangle, \label{Ai}
\end{align}

\begin{align}
   B_{ij}({\bf p}) =& \frac{1}{2 E_p  \gamma_{HQ}} \int \frac{d^3 {\bf q}}{(2 \pi)^3 E_q}  \int \frac{d^3 {\bf q'}}{(2 \pi)^3 E_{q'}} \int \frac{d^3 {\bf p'}}{(2 \pi)^3 E_{p'}}\nonumber\\
   &\times\sum{|{\mathcal{M}}_{2\rightarrow2}|^2} \ (2 \pi)^4 \  \delta^{(4)}(p+q-p'-q') \nonumber\\
   & \times f_k(E_q) \ (1\pm f_k(E_{q'})) \ [(p-p')_i]\nonumber\\
   =& \frac{1}{2} \langle\langle (p-p')_i \ ( p - p')_j \rangle\rangle, \label{Bij}
\end{align}
where $\gamma_{HQ}=N_s \times N_c$ is the heavy quark degeneracy factor with $N_s=2$, $N_f=3$ ($u,d,s$), and $N_c=3$ (for $SU(3)$).
$|{\mathcal{M}}_{2\rightarrow2}|$ represents the scattering amplitude  for $2\rightarrow2$ process (refer Appendix \textbf{A} of ref.~\cite{Shaikh:2021lka}) and $p=|{\bf p}|$ is the magnitude of heavy quark initial momentum. $f_k(E_q)$ denotes the distribution function for quarks, antiquarks ($k=lq, l\bar{q}$) and gluons ($k=g$). For the final state phase space, we have the Fermi suppression factor $(1 - f_{lq}(E_{q'}))$ for light quarks and Bose enhancement factor $(1 + f_{g}(E_{q'}))$ for gluons. Here, the drag force gives the thermal average of the momentum transfer between the initial and final states of the bottom quark, whereas $B_{ij}$ denotes the square of the momentum transfer as the bottom quark diffuses through the medium. Since, both $A_i({\bf p})$ and $B_{ij}({\bf p})$ depend explicitly on the initial heavy quark momentum $({\bf p})$, the drag coefficient ($A$) is defined as,
\begin{align}
    A_i =& \ p_i \ A(p^2) \implies A(p^2)=\langle\langle 1 \rangle\rangle - \frac{\langle\langle {\bf{p.p'} \rangle\rangle}}{p^2},\label{Ap}.
\end{align}
The diffusion tensor decomposed into its transverse and longitudinal components give,
\begin{align}
    B_{ij} =& \left[ \delta_{ij} - \frac{p_ip_j}{p^2} \right]B_0(p^2) + \frac{p_ip_j}{p^2} B_1(p^2),\label{Bijp} 
\end{align}
where, the transverse momentum diffusion ($B_0$) and longitudinal momentum diffusion ($B_1$) coefficients are defined as, 
\begin{align}
&B_{0}= \frac{1}{4}\left[\langle\langle p'^{2} \rangle\rangle-\frac{\langle\langle ({\bf{p.p'}})^2\rangle\rangle}{p^2} \right],\label{B0p}\\ 
&B_{1}= \frac{1}{2}\left[\frac{\langle\langle ({\bf{p.p'})}^2\rangle\rangle}{p^2} -2\langle\langle ({\bf{p.p'})}\rangle\rangle +p^2 \langle\langle 1 \rangle\rangle\right]\label{B1p},
\end{align}\\

 In the center-of-momentum frame of the colliding system, the thermal average of a function $F({\bf p})$ for $2\rightarrow 2$ process, in general, becomes,
\begin{align}
    \langle \langle F({\bf p})\rangle \rangle_{col}&=\frac{1}{(512 \, \pi^4) E_p \gamma_{HQ}}\int_0^\infty dq \left(\frac{s-m_{HQ}^2}{s}\right) f_k(E_q)\nonumber \\ &\times (1\pm f_k(E_{q'})) \int_0^\pi d\chi \, \sin\chi \int_0^\pi d\theta_{cm} \, \sin\theta_{cm} \nonumber\\ &\times\sum{|{\mathcal{M}}_{2\rightarrow2}|^2} \int_0^{2\pi} d\phi_{cm} \ F({\bf p}),\label{Fcm}
\end{align}

where $\chi$ is the angle between the heavy quark and medium particles in the lab frame and $s=(E_p+E_q)^2-|{\bf p}|^2-|{\bf q}|^2-2|{\bf p}||{\bf q}| \cos \chi$. The zenith $\theta_{cm}$ and azimuthal $\phi_{cm}$ angles are defined in the center-of-momentum frame. The Debye screening mass ($m_D$) is inserted at leading order for $t$-channel gluonic propagator to the in-medium matrix elements,
\begin{equation}
    m_D^2=(4\pi\alpha_s)\,T^2\left(\frac{N_c}{3}+\frac{N_f}{6}\right).
\end{equation}

\subsubsection{\textbf{\textit{2$\rightarrow$3 process}}}
%
Heavy quarks can also radiate soft gluons while moving through the QGP medium along with the collisions. The inelastic ($2\rightarrow 3$) process is,
\begin{equation}
    HQ \ (p)+lq/l\bar{q}/g \ (q) \rightarrow HQ \ (p')+lq/l\bar{q}/g  \ (q') + g \ (k'),\label{Rad}
\end{equation}  
where $k'\equiv(E_{k'},{\bf k'_{\perp}},k'_z)$ is the four-momentum of the final state soft gluon emitted by the bottom quark ($k'\rightarrow 0$). Compared to the collisional process, for the radiative process, only the kinematical and the interaction parts are modified in Eqs. (\ref{Ai}) and (\ref{Bij}).
The general expression for the thermal averaged $F({\bf p})$ for $2\rightarrow 3$ process is ~\cite{Mazumder:2013oaa},
\begin{align}
    \langle \langle F({\bf p}) &\rangle \rangle_{rad} =\frac{1}{2 E_p  \gamma_{HQ}} \int \frac{d^3 {\bf q}}{(2 \pi)^3 E_q}  \int \frac{d^3 {\bf q'}}{(2 \pi)^3 E_{q'}} \int \frac{d^3 {\bf p'}}{(2 \pi)^3 E_{p'}} \nonumber\\ &\times\int \frac{d^3 {\bf k'}}{(2 \pi)^3 E_{k'}} \sum{|{\mathcal{M}}_{2\rightarrow 3}|^2} \  \delta^{(4)}(p+q-p'-q'-k') \nonumber\\&\times (2 \pi)^4 \ f_k(E_q) \ (1 \pm f_k(E_{q'})) \ (1 + f_k(E_{k'})) \ \nonumber\\& \times \theta_1(E_p-E_{k'}) \ \theta_2(\tau-\tau_F) \ F({\bf p}),\label{Frad}
\end{align}
where the theta function $\theta_1(E_p-E_{k'})$ ensures that the bottom quark initial energy $E_p$ is always greater than radiated soft gluon energy $E_{k'}$ and $\theta_2(\tau-\tau_F)$ keep the collision time $\tau$ of the heavy quark with the medium particles always greater than the gluon formation time $\tau_F$ (Landau-Pomeranchuk-Migdal Effect)~\cite{Wang:1994fx,Gyulassy:1993hr,Klein:1998du}. $(1 + f_g(E_{k'}))$ is the Bose enhancement factor for the radiated gluon and $|{\mathcal{M}}_{2\rightarrow 3}|^2$ is the matrix element squared for the radiative process which can be written as \cite{Abir:2011jb},
\begin{equation}\label{M23}
    |{\mathcal{M}}_{2\rightarrow 3}|^2=|{\mathcal{M}}_{2\rightarrow 2}|^2 \times  \frac{12 g_s^2}{(k'_\perp)^2} \left(1+\frac{m_{HQ}^2}{s}e^{2y_{k'}}\right)^{-2}, 
\end{equation}
where $y_{k'}$ is the rapidity of the emitted gluon and 
$\left(1+\frac{m_{HQ}^2}{s}e^{2y_{k'}}\right)^{-2}$ is the dead cone factor for the heavy quark. The details of the soft gluon 3-momentum integral are discussed in detail in Appendix {\bf B} of ref. \cite{Shaikh:2021lka}.

\subsection{EQPM distribution of quarks and gluons in a viscous medium}
%
EQPM include the effects of the thermal interactions of the viscous QGP medium in the analysis via a realistic QCD equation of state. For the  system close to the local equilibrium, the in-medium particle momentum distribution function has the form, 
\begin{align}\label{2.1}
 &f_k=f^0_k+\delta f_k \hspace{0.75 cm} \text{with}  &&\delta f_k/f^0_k \ll1,
\end{align} 
where $f^0_k$ is the EQPM equilibrium distribution function and $\delta f_k$ is the non-equilibrium component. The EQPM distribution functions of light quarks/antiquarks and gluons are defined in terms of effective fugacity parameter $z_k$ which encodes the QCD medium interactions as follows (for zero baryon chemical potential),
\begin{align}\label{2.2}
&f^0_{lq/l\bar{q}} =\frac{z_{lq/l\bar{q}} \exp[-\beta (u\!\cdot\! q)]}{1 + z_{lq/l\bar{q}}\exp[-\beta (u\!\cdot\! q)]},\\
&f^0_g =\frac{z_g \exp[-\beta\, (u\!\cdot\! q)]}{1 - z_g\exp[-\beta\, (u\!\cdot\! q)]}.\label{2.202}
\end{align}

The parameters $z_{lq/l\bar{q}}$ and $z_g$ are effective fugacities that encode the QCD interactions for quarks/antiquarks and gluons. These temperature-dependent parameters modify the single-particle dispersion relation as,
\begin{equation}\label{2.3}
\Tilde{q_k}^{\mu} = q_k^{\mu}+\delta\omega_k\, u^{\mu}, \qquad
\delta\omega_k= T^{2}\,\partial_{T} \ln(z_{k}),
\end{equation}
where $\Tilde{q}_k^{\mu}=(\omega_k, {\bf q}_k)$ and $q_k^{\mu}=(E_q, {\bf q}_k)$ are the quasiparticle (dressed) and bare particle momenta, respectively. In the limit, $z_k\rightarrow 1$, the ideal equation of state is reproduced.\\ 

The effective coupling constant ($\alpha_{eff}$) is introduced from the kinetic theory through EQPM-based Debye mass as~\cite{Mitra:2017sjo},
\begin{equation}\label{2.4}
\frac{\alpha_{eff}}{\alpha_s(T)}= \dfrac{\frac{2N_c}{\pi^2}\mathrm{PolyLog}~[3,z_g]
-\frac{2N_f}{\pi^2}\mathrm{PolyLog}~[3,-z_{lq}]}{\left(\frac{N_c}{3}+\frac{N_f}{6}\right)}.
\end{equation}
$\alpha_{s}(T)$ is the 2-loop running coupling constant at finite temperature ~\cite{Kaczmarek:2005ui,Das:2015ana}.

The evolution of the medium particle distribution function can be described by the effective Boltzmann equation based on the EQPM and using the effective covariant kinetic theory approach. Within RTA, it is expressed as follows~\cite{Mitra:2018akk},
\begin{equation}\label{2.6}
\Tilde{q}^{\mu}_k\,\partial_{\mu}f_k(x,\Tilde{q}_k)+F_k^{\mu}\left(u\!\cdot\!\tilde{q}_k\right)\partial^{(q)}_{\mu} f_k = -\left(u\!\cdot\!\tilde{q}_k\right)\frac{\delta f_k}{\tau_R},
\end{equation}
where $\tau_{R}$ is the thermal relaxation time and $F_k^{\mu}=-\partial_{\nu}(\delta\omega_k u^{\nu}u^{\mu})$ is the mean field force term. The viscous corrections to the distribution function are obtained by solving Eq.~(\ref{2.6}) using the iterative Chapman-Enskog  method~\cite{Jaiswal:2013npa} and obtain the first order correction to the distribution function as,
\begin{align}\label{2.7}
\!\!\delta f_k = \tau_R\bigg( \Tilde{q}_k^\gamma\partial_\gamma \beta + \frac{\beta\, \Tilde{q}_k^\gamma\, \Tilde{q}_k^\phi}{u\!\cdot\!\Tilde{q}_k}\partial_\gamma u_\phi -\beta\theta\,\delta\omega_k \bigg) f^0_k\Tilde{f}^0_k, 
\end{align}
where $\tilde{f_k}^0\equiv(1-a_kf_k^0)$ ($a_{g}=-1$ for bosons and $a_{lq}=+1$ for fermions). The first-order evolution equation for the shear stress tensor $\pi^{\mu\nu}$ and bulk viscous pressure $\Pi$ within the effective kinetic theory has the following forms~\cite{Bhadury:2019xdf},
\begin{align}\label{2.8}
\pi^{\mu\nu} = 2\,\tau_R\,\beta_\pi\,\sigma^{\mu\nu}, \ \ \ \ \ 
\Pi = -\tau_R\,\beta_\Pi\,\theta, 
\end{align}
with $\theta\equiv\partial_{\mu}u^{\mu}$ as the scalar expansion parameter and $\sigma^{\mu\nu}\equiv\Delta^{\mu\nu}_{\alpha\beta}\nabla^{\alpha}u^{\beta}$ where $\Delta^{\mu\nu}_{\alpha\beta}\equiv\frac{1}{2}(\Delta^\mu_\alpha\Delta^\nu_\beta +\Delta^\mu_\beta\Delta^\nu_\alpha)-\frac{1}{3}\Delta^{\mu\nu}\Delta_{\alpha\beta}$ denotes  traceless symmetric projection operator orthogonal to the fluid velocity $u^{\mu}$.
Here, $\beta_\pi$ and $\beta_\Pi$ are the first-order coefficients and have the form specified in ref. ~\cite{Bhadury:2019xdf}.

Using the evolution equation, the shear and bulk viscous corrections to the distribution function can be expressed as, 
\begin{align}
\delta f_k&=\delta f_k^{\text{shear}}+\delta f_k^{\text{bulk}},\\
\delta f_k^{\text{shear}}&=\dfrac{\beta\,f_k^0{\tilde{f}}^0_k}{2\beta_{\pi}(u.\Tilde{q}_k)}\Tilde{q}_k^{\alpha}\Tilde{q}_k^{\beta}\pi_{\alpha\beta},\label{2.11}\\ 
\delta f_k^{\text{bulk}}&=-\dfrac{\beta\,f_k^0{\tilde{f}}^0_k}{\beta_{\Pi}(u.\Tilde{q}_k)}\Big[(u.\Tilde{q}_k)^2c_s^2-\frac{\mid{\Tilde{\bf q}}_k\mid^2}{3}-(u.\Tilde{q}_k)\delta\omega_k\Big]\Pi.\label{2.12}
\end{align}
Substituting the modified in-medium particle distribution functions with the shear and bulk viscous corrections from Eq.~(\ref{2.11}) and~(\ref{2.12}) into  Eqs.~(\ref{Ap})-(\ref{B1p}), we obtain the modified heavy quark drag and diffusion coefficients in the viscous medium up to the first order. Considering longitudinal boost invariant expansion through Bjorken prescription~\cite{Bjorken:1982qr} using Milne coordinates $(\tau,x,y,\eta_s)$, the Eq.~(\ref{2.11}) and Eq.~(\ref{2.12}) simplifies to,
\begin{align}
    \delta f_k^{\text{shear}}&=\frac{f_k^0{\tilde{f}}^0_k\,\mathfrak{s}}{\beta_\pi \omega_k T \tau} \left(\frac{\eta}{\mathfrak{s}}\right) \left[\frac{|{\bf q_k}|^2}{3}-(q_k)_z^2\right],\label{phishearmod}\\
    \delta f_k^{\text{bulk}}&=\frac{f_k^0{\tilde{f}}^0_k\,\mathfrak{s}}{\beta_\Pi \omega_k T \tau} \left(\frac{\zeta}{\mathfrak{s}}\right) \left[(\omega_k)^2 c_s^2 - \frac{|{\bf q_k}|^2}{3} - (\omega_k)\delta\omega_k \right],\label{phibulkmod}
\end{align}
where $\tau=\sqrt{t^2-z^2}$ is the proper time and $\eta_s=\tanh^{-1}(z/t)$ is the space-time rapidity
 with $u^\mu=(1, 0, 0, 0)$ and $g^{\mu\nu}=(1,-1,-1,-1/\tau^2)$. Here, $\theta=1/\tau$, $\Pi=-\zeta/\tau$ and $\pi^{\mu\nu}\sigma_{\mu\nu}=4\eta/3\tau^2$. $\eta$ and $\zeta$ denote the shear and bulk viscosity of the QGP respectively, $c_s^2$ is the speed of sound squared and $\mathfrak{s}$ is the entropy density of the medium.\\

%
\section{Results and discussions}\label{Results}
%

\begin{figure*}
\includegraphics[scale=0.55]{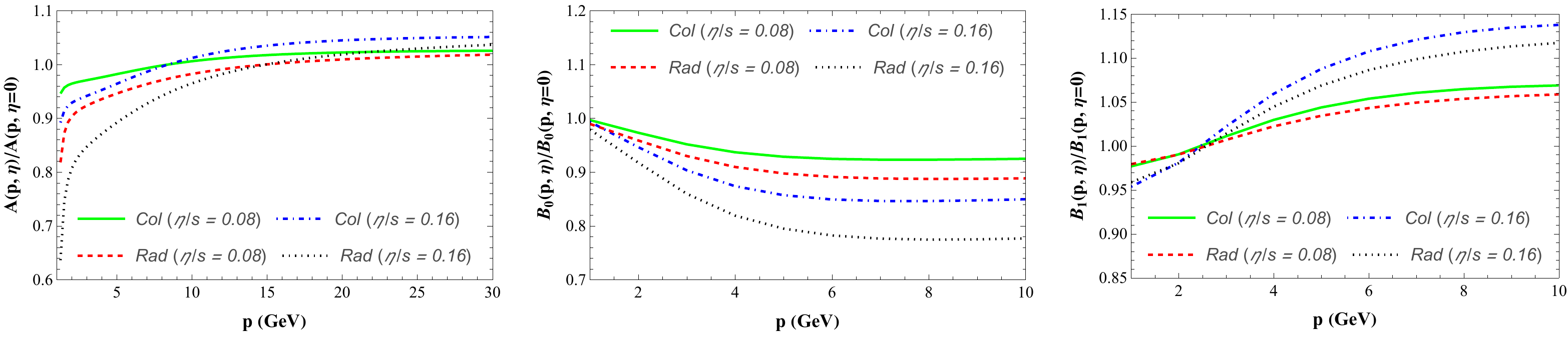} 
\caption{\small Bottom quark transport coefficients with first-order shear viscous correction scaled with the corresponding value for the non-viscous case $(\eta=0)$ as a function of its initial momentum ($p$) at $T=3\,T_c$}
\label{fig:1}
\end{figure*}

\begin{figure*}
\includegraphics[scale=0.55]{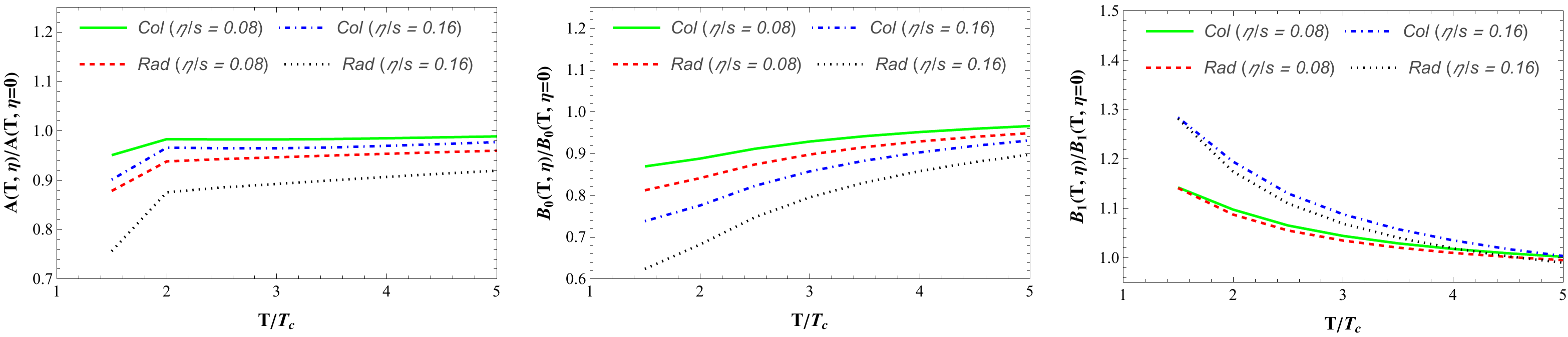} 
\caption{\small Bottom quark transport coefficients with first-order shear viscous correction scaled with the corresponding value for the non-viscous case $(\eta=0)$ as a function of QGP temperature ($T/T_c$) for the initial momentum $p=5$ GeV.}
\label{fig:2}
\end{figure*}

\begin{figure*}
\includegraphics[scale=0.7]{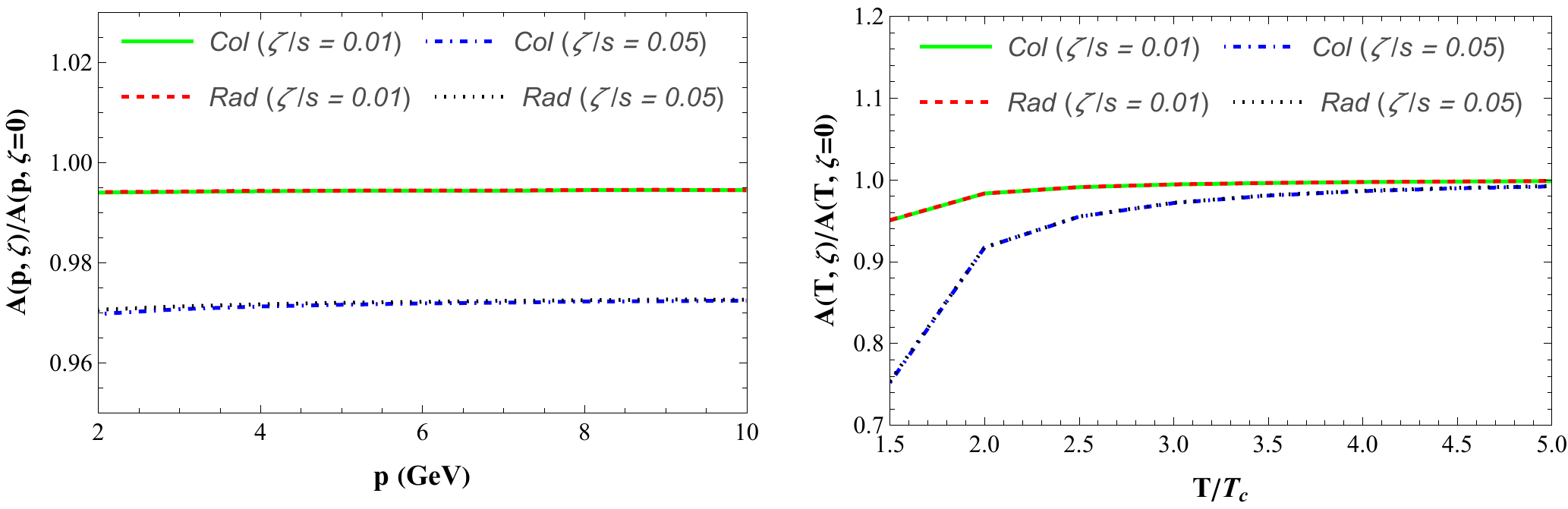} 
\caption{\small Bottom quark drag coefficient with first-order bulk viscous correction and scaled with the corresponding value for the non-viscous case $(\zeta=0)$ as a function of its initial momentum ($p$) at $T=3\,T_c$ (left panel), and as a function of scaled QGP temperature ($T/T_c$) at $p=5$ GeV (right panel).}
\label{fig:3}
\end{figure*}

\subsection{Bottom transport coefficients with shear viscous correction}

In this study, we take the bottom quark mass $m_b=4.2$ GeV, quark-hadron transition temperature $T_c = 170$ MeV for three massless light quark flavors with zero net baryon density and the proper time $\tau = 0.25$ fm. Fig.\,\ref{fig:1} displays the effects of shear viscous corrections on the momentum dependence of the bottom quark drag coefficients for collisional and radiative processes in the QGP. The transport coefficients are scaled with their respective values for the $\eta=0$ case. Both the momentum and temperature dependence of the transport coefficients due to the collisional and radiative processes can be described using Eqs. (\ref{Ap}-\ref{Fcm}) and (\ref{Frad}). It is observed that the shear viscosity substantially reduces the heavy quark drag ($A(\eta)/A(\eta=0)$) (left panel) at low momenta ($p\approx 1-8$ GeV for collision and $p\approx 1-14$ GeV for radiation). However, in the high momentum regime ($p \gtrsim 15$ GeV), the drag coefficient increases with an increase in the shear viscosity to entropy density ratio $\eta/\mathfrak{s}$. This can be understood from the interplay of two terms in Eq.~(\ref{Ap}) in different momentum regimes while incorporating the viscous effects through Eq.~(\ref{phishearmod}). The shear viscous correction is more prominent at the low momenta of the bottom quark, with an increase in the shear viscosity resulting in a decrease in the drag coefficient for both collisional and radiative processes. The radiative curves for low initial momentum seem to be affected more compared to the collisional ones whereas as we look towards the higher momentum side $p\approx 30$ GeV the radiative curves show less effect to the change in $\eta/\mathfrak{s}$ since the radiative process is suppressed for bottom quark due to the dead-cone effect for $\theta_k \sim m_{HQ}/E_{HQ}$. In the Fig.\,\ref{fig:1} (middle panel), the scaled transverse diffusion coefficient $B_0(\eta)/B_0(\eta=0)$ deviate slightly from the ideal case ($\eta=0$) for low momentum with subsequent suppression towards higher momentum values. Further, the increase in shear viscosity leads to more deviation from the equilibrium with the radiative curves being more suppressed than the collision ones.
The longitudinal diffusion coefficient ($B_1(\eta)/B_1(\eta=0)$) (right panel) is reduced due to shear viscous correction at low momenta ($p\lesssim2$ GeV) affecting both collisional and radiative curves equally with the variation of $\eta/\mathfrak{s}$. For momenta $p\gtrsim4$ GeV, the behaviour is observed to be quite the opposite for low momenta wherein the longitudinal diffusion coefficient increases as compared to its value in the absence of shear viscosity, with higher $\eta/\mathfrak{s}$ resulting in a larger deviation.

Fig. \ref{fig:2} shows the effect of variation of the transport coefficients of the bottom quark as a function of the scaled QGP temperature ($T/Tc$). Introducing shear viscous effects considerably reduces the bottom quark drag coefficient near $T_c$. This behaviour can be accounted by the negative contribution from the factor $\left[\frac{|{\bf q_k}|^2}{3}-(q_k)_z^2\right]$ in  Eq.~(\ref{phishearmod}) for $\delta f_k$. The shear viscous effect is more pronounced near the transition temperature due to the temperature dependence of $\beta_\pi$ ($\beta_{\pi}\propto T^4$) such that $\frac{\mathfrak{s}}{\beta_\pi T}\propto \frac{1}{T^2}$ in Eq.~(\ref{phishearmod}). Qualitatively similar behaviour is observed for the temperature dependence of transverse diffusion coefficient $B_0(\eta)/B_0(\eta=0)$ (middle panel) of the bottom quark with suppression of the ratio near $1.5 T_c$. There is also an overall suppression with increasing $\eta/\mathfrak{s}$ for both collisional and radiative curves, with the latter being affected more over the entire temperature regime considered here. Contrary to the drag and transverse diffusion coefficients described earlier, the longitudinal momentum diffusion coefficient $B_1(\eta)/B_1(\eta=0)$ (right panel) for the bottom quark momentum of $p=5$ GeV seems to increase with an increase in $\eta/\mathfrak{s}$ for both collision and radiative processes for $T<4\,T_c$. Following the same arguments for the temperature behaviour of heavy quark drag coefficient, the shear viscous effect (entering through the $\delta f_k$ with $\frac{\mathfrak{s}}{\beta_\pi T}\propto \frac{1}{T^2}$) to the diffusion coefficients is more visible in the low-temperature regimes.

\subsection{Bottom transport coefficients with bulk viscous correction}
In Fig. \ref{fig:3} (left panel), the effect of bulk viscous correction on the momentum dependence of the bottom quark drag coefficient ($A(\zeta)/A(\zeta=0)$) is shown. The bulk viscosity seems to reduce the heavy quark drag with almost constant dependence throughout the momentum range considered here. For both collisional and radiative energy loss, the drag coefficient ratio decreases with an increase in bulk viscosity to entropy density ratio ($\zeta/\mathfrak{s}$). This can be understood by the suppression of $\delta f_k^{\text{bulk}}$ due to the negative terms in Eq.~(\ref{phibulkmod}) with an increase in $\zeta/\mathfrak{s}$. Fig.\,\ref{fig:3} (right panel) shows the effect of the variation of the scaled drag coefficient as a function of the scaled QGP temperature $T/T_c$. The bulk viscosity effect is prominent near the transition temperature $T\approx1.5\,T_c$ and the drag coefficient approaches the conformal limit ($c_s^2\approx\frac{1}{3}$) at high temperatures $T\gtrsim4\,T_c$ . This behaviour can be explained from Eq.~(\ref{phibulkmod}) wherein for $T>>T_c$, we have $z_k\rightarrow 1$ (ideal equation of state) and the medium modified part of the dispersion relation vanishes, $i.e.$, $\delta\omega_k\rightarrow 0$ in Eq.~(\ref{2.3}). The remaining two terms get cancelled for our case of the massless quasiparticles, leading to effectively zero contribution from $\delta f_k^{\text{bulk}}$ at high temperature. We also notice that both collisional and radiative processes show identical behaviour with the variation in bulk viscosity; consistent with our previous study for charm quark ~\cite{Shaikh:2021els}. The transverse and longitudinal diffusion coefficients plots including the bulk viscous corrections show similar trends as for the drag coefficient case shown in fig. \ref{fig:3}.

\begin{figure*}
\includegraphics[scale=0.7]{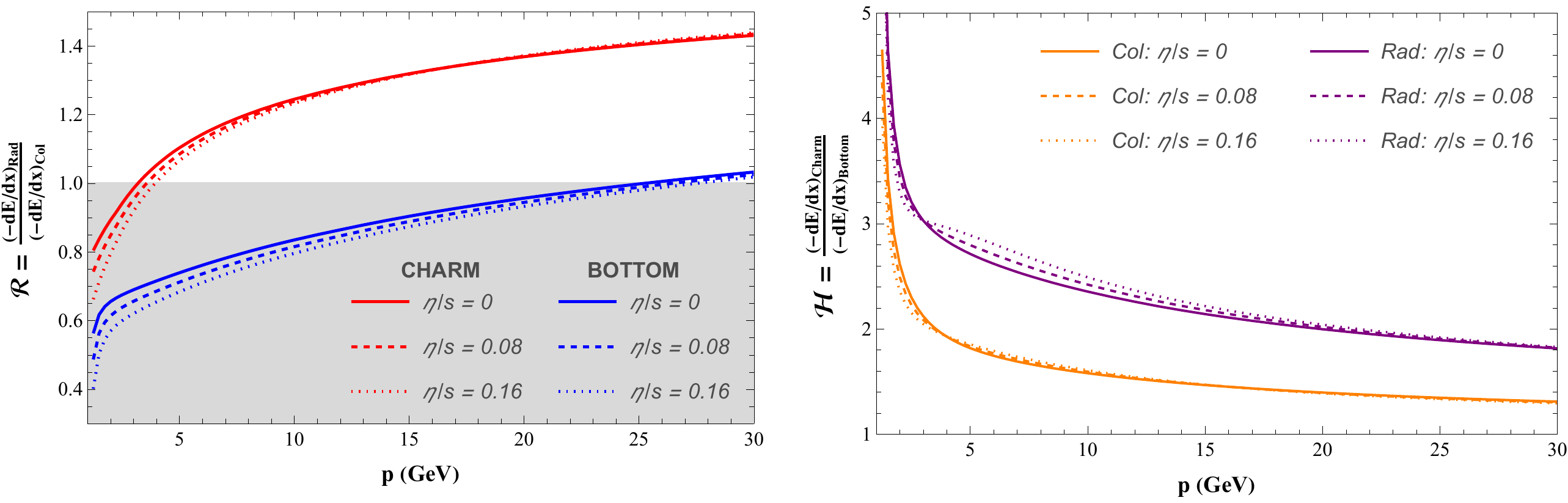} 
\caption{\small (Left panel) Ratio of the radiative to collisional differential energy loss for the charm quark (red curves) and the bottom quark (blue curves) at $3\,T_c$. (Right panel) Ratio of charm to bottom quarks energy loss due to collision (orange curves) and radiation (purple curves) at $3\,T_c$.}
\label{fig:4}
\end{figure*}

\subsection{Collisional and radiative energy loss of charm and bottom quarks}

The drag coefficient corresponds to the resistance to the motion of the heavy quark by the thermal QGP constituents. The differential energy loss of the heavy quark is related to its drag coefficient as \cite{GolamMustafa:1997id},
\begin{equation}\label{EL}
    -\frac{dE}{dx} = p\,A(p).
\end{equation}
Fig. \ref{fig:4} (left panel) shows the ratio ($\mathcal{R}$) of the differential energy loss for the radiative (inelastic) process and compared with the collisional (elastic) energy loss for the charm ($m_c=1.3$ GeV) and bottom quarks ($m_b=4.2$ GeV) for different values of shear viscosity to entropy density ratio ($\eta/\mathfrak{s}$) in the viscous QCD medium at $T=3\,T_c=510$ MeV. For both charm and bottom, increasing shear viscosity decreases the radiative to collisional energy loss ratio near low momentum $p\sim m_{HQ}$. However, the viscous corrections have a negligible effect at high momentum ($p >> m_{HQ}$).  The suppression in the radiative energy loss of the bottom quark is significantly more compared to the charm quark.  The radiative dominance for the bottom quark occurs at around $30$ GeV which is almost ten times larger than that for the charm quark where the ratio exceeds one at around $3$ GeV. This behaviour can be attributed to the dead-cone effect which suppresses the soft-gluon radiation by a heavy quark at small angles. The bottom quark is roughly three times more massive compared to the charm quark. Hence, its dead cone angle $\theta_k\sim m_{HQ}/E_{HQ}$, where there is the absence of gluon radiation is larger and the probability of energy loss due to radiation is lesser compared to the charm. Also, notice that the slope of the ratio where the radiative dominance occurs is larger for charm compared to the bottom.\\

Fig. \ref{fig:4} (right panel) shows the ratio ($\mathcal{H}$) of the differential energy loss for the charm quark ($m_c=1.3$ GeV) compared with the bottom quark ($m_b=4.2 $ GeV) for collision and radiative processes at different values of specific shear viscosity ($\eta/\mathfrak{s}$). This ratio is equivalent to the ratio of the drag coefficients of the charm to the bottom quarks (from eq. (~\ref{EL})). Two interesting features can be observed in this plot. First, this ratio ($\mathcal{H}$) is greater for the radiative process implying that the charm quark suffers more energy loss due to radiation than the bottom quark. Also, the ratio for the collisional process is closer to one compared to radiation implying the difference in energy loss mechanisms between charm and bottom owing to their dissimilar mass can be attributed significantly to the soft gluon radiation. Secondly, the effect of the change in $\eta/\mathfrak{s}$ over the range $0$ to $0.16$ seems prominent for the radiative process between $2$ GeV to $15$ GeV compared to collision. This suggests that the medium-induced bremsstrahlung of heavy quarks is affected more due to non-equilibrium effects generated through the viscosity of the medium and it is larger for the charm quark in comparison to the bottom.

%
\section{Conclusion}\label{Conclusion}
%

The bottom quark transport coefficients have been investigated in the viscous QGP by considering its Brownian motion in the medium using the Fokker-Planck dynamics. The energy loss in the inelastic process of soft gluon radiation by the bottom quark is studied along with the elastic collision with the medium constituents. The thermal medium interactions are included through EQPM in the analysis through the temperature-dependent effective fugacity parameter for the in-medium particles. The shear and bulk viscous corrections at leading order are incorporated into the quarks, antiquarks, and gluon momentum distribution functions which are obtained by solving the effective Boltzmann equation within the EQPM framework.\\

The shear and bulk viscous corrections to the bottom quark drag and momentum diffusion coefficients have been estimated as a function of its initial momentum and the QGP temperature. We observe that the shear viscous correction to the bottom quark transport coefficients is prominent at the low initial momentum of the bottom quark and near the transition temperature $T_c$. Similar results for the charm quark have been observed in an earlier study ~\cite{Shaikh:2021lka}. The radiative curves seem to be affected slightly more due to a change in $\eta/\mathfrak{s}$ compared to the collision case.\\

Results obtained by including the first-order bulk viscous correction are also presented here for the bottom quark drag coefficient. Bulk viscous correction tends to decrease the heavy quark transport coefficients consistently throughout the entire momentum regime considered here with further suppression due to increasing $\zeta/\mathfrak{s}$. The temperature dependence of the bottom transport coefficient due to bulk correction shows a considerable modification of the transport coefficient ratio near $T_c$ which at the high temperature approaches the conformal limit ($\epsilon=3P$). The effect of including bulk viscous correction is almost identical for the collisional and radiative processes, suggesting its minimal effect on the two processes compared to the shear viscous correction.\\

The differential energy loss ratios for the radiative to collisional process and the charm to bottom quark energy loss have been investigated and comparatively studied. We conclude that radiation is the dominant process of energy loss for charm quark with momentum $p\gtrsim 4 $ GeV. However, for the bottom quark, the radiative energy loss becomes significant at a much higher value of its initial momentum, indicating the importance of considering the radiative process at momenta $p\gtrsim 30$ GeV.

\begin{acknowledgements}
The authors acknowledge Santosh Kumar Das for reading and providing useful comments on the manuscript.
\end{acknowledgements}

\appendix
%
\section{Thermodynamic integrals}\label{A3}
%
The thermodynamic integrals $\Tilde{J}^{(r)}_{k~nm}$ and $\Tilde{L}^{(r)}_{k~nm}$ are respectively defined as follows,
\begin{align}
   \Tilde{J}^{(r)}_{k~nm}&=\frac{\gamma_k}{2\pi^2}\frac{(-1)^m}{(2m+1)!!}\int_{0}^\infty{d\mid{\Tilde{\bf p}}_k\mid}~\big(u\cdot\Tilde{p}_k \big)^{n-2m-r-1}\nonumber\\
&\times\big(\mid{\Tilde{\bf p}}_k\mid\big)^{2m+2} f^{0}_k\, \tilde{f}^0_k,\\
\Tilde{L}^{(r)}_{k~nm}&=\frac{\gamma_k}{2\pi^2}\frac{(-1)^m}{(2m+1)!!}\int_{0}^\infty{d\mid{\Tilde{\bf p}}_k\mid}~\frac{\big(u.\Tilde{p}_k\big)^{n-2m-r-1}}{\mid{\Tilde{\bf p}}_k\mid}\nonumber\\
    &\times\big(\mid{\Tilde{\bf p}}_k\mid\big)^{2m+2}f^0_k\tilde{f}^0_k. 
\end{align}
For the massless limit of the light quark, the thermodynamic integrals can be expressed in terms of the $PolyLog$ function as follows,
\begin{align}
    \Tilde{J}^{(1)}_{k~42} =&-\frac{2\,a_k\gamma_kT^5}{5\pi^2}\bigg[2\,\mathrm{PolyLog}~[4,-a_kz_k]\nonumber\\&
    -\frac{\delta\omega_k}{T}\,\mathrm{PolyLog}~[3,-a_kz_k]\bigg],
    \end{align}
\begin{align}
    \Tilde{L}^{(1)}_{k~42} &=-\frac{a_k\gamma_kT^4}{5\pi^2}~\mathrm{PolyLog}~[3,-a_kz_k].
\end{align}

\bibliography{reference}{}

\end{document}